\documentclass[aps,]{revtex4}
\usepackage{graphicx}
\usepackage{subfigure}

\newcommand{\bb}{\beta}
\newcommand{\bc}{\beta_c}

\newcommand{\te}{\tau_{\rm exp}}

\newcommand{\ps}{p_{\rm swap}}

\newcommand{\pp}{c}
\newcommand{\s}{\sigma}
\newcommand{\N}{K}
\newcommand{\sig}{\Delta}
\newcommand{\W}{{\cal W}}

\newcommand{\g}{j}

\newcommand{\wn}{\rho}
\newcommand{\F}{{\tilde F}}

\newcommand{\wt}{Q}

\usepackage{amsmath,amssymb,graphicx}
\usepackage{algorithmic}

\newcommand{\T}{S}
\newcommand{\bp}{\beta^\prime}
\newcommand{\sr}{\s}
\newcommand{\ppt}{{\tilde c}}

\begin{document}

\title{Monte Carlo Methods for Rough Free Energy Landscapes: Population Annealing and Parallel Tempering
}

\author{J.~Machta}
\email{machta@physics.umass.edu}
\affiliation{
Physics Department,
University of Massachusetts,
Amherst, MA 01003 USA}

\author{R. S.~Ellis}
\email{rsellis@math.umass.edu}
\affiliation{
Department of Mathematics and Statistics,
University of Massachusetts,
Amherst, MA 01003 USA}

\begin{abstract}
Parallel tempering and population annealing are both effective methods for simulating equilibrium systems with rough free energy landscapes. Parallel tempering, also known as replica exchange Monte Carlo, is a Markov chain Monte Carlo method while population annealing is a sequential Monte Carlo method.  Both methods overcome the exponential slowing associated with high free energy barriers.  The convergence properties and efficiency of the two methods are compared. For large systems, population annealing initially converges to equilibrium more rapidly than parallel tempering for  the same amount of computational work.  However, parallel tempering converges exponentially  and population annealing inversely in the computational work so that ultimately parallel tempering approaches equilibrium more rapidly than population annealing.
\end{abstract}
\maketitle
\section{Introduction}
\label{sec:intro}

Equilibrium systems with rough free energy landscapes, such as spin glasses, configurational glasses and proteins, are difficult to simulate using conventional Monte Carlo methods because the simulation tends to be trapped in metastable states and fails to explore the full configuration space.  A number of techniques have been proposed to overcome this problem.  Some of these techniques involve simulating an extended state space that includes many temperatures~\cite{Okamoto04, NaHa07a}.  Multicanonical simulations, umbrella sampling, the Wang Landau method, simulated tempering and parallel tempering all fall into this class.  Parallel tempering~\cite{SwWa86,Geyer91,HuNe96, EaDe05}, also known as replica exchange Monte Carlo, is perhaps the most widely used of these methods because it is simple to program and performs well in many settings.   It is the standard method for simulating spin glasses~\cite{KaKoYo06,BaCrFe10} and is used for protein folding~\cite{Ha97,ScHeVeWe05} and lattice gauge theory~\cite{BuFuKeMu07}.  

Parallel tempering and the other members of its class are all Markov chain Monte Carlo methods.  In Markov chain Monte Carlo, the target distribution is approached via repeated application of an elementary process that typically satisfies detailed balance with respect to the target distribution.  In the case of parallel tempering, the target distribution is a joint distribution whose marginals are equilibrium distributions for a set of temperatures.  In sequential Monte Carlo, by contrast,  the target distribution is the last member of a sequence of distributions, each of which is visited once.   The initial distribution is easy to equilibrate and a resampling step transforms one distribution to the next in the sequence. Population annealing~\cite{HuIb03, Mac10a} is a sequential Monte Carlo algorithm in which the sequence of distributions are equilibrium distributions of decreasing temperature.   

In this paper we will describe both parallel tempering and population annealing and compare their efficiency and convergence properties in the context of a simple, tractable free energy landscape comprised of two wells separated by a high barrier.  Although this free energy landscape is highly simplified compared to the landscapes of more realistic models, we believe that it captures some of the essential features of rough free energy landscapes and that the lessons learned from this analysis will be useful in understanding and improving the performance of both parallel tempering and population annealing in realistic settings. In Ref.\ \cite{Mac09a} we analyzed the performance of parallel tempering for this landscape. 

Although they are based on quite different strategies, parallel tempering (PT) and population annealing (PA) share a number of common features. Both are methods that build on a conventional Markov chain Monte Carlo procedure whose stationary distribution is a fixed temperature equilibrium ensemble,  such as the Metropolis algorithm or Glauber dynamics.  We refer to this procedure as the {\em equilibrating subroutine}.  At sufficiently high temperature the equilibrating subroutine converges rapidly to the equilibrium ensemble.  Both PT and PA take advantage of this rapid  equilibration at high temperature to accelerate the convergence to equilibrium at lower temperatures.  Both PT and PA attempt to transform equilibrium high temperature configurations into equilibrium low temperature configurations through a sequence of temperature steps such that the system remains close to equilibrium.   In PT there is a single replica of the system at each temperature in the sequence, and replicas are allowed to move between temperatures via replica exchange.  These replica exchange moves are carried out with acceptance probabilities that satisfy detailed balance so that the entire set of replicas tends toward equilibrium at their respective temperatures.  

Population annealing is closely related to simulated annealing~\cite{KiGeVe83,LaAa87}.  In simulated annealing a single realization of the system is cooled from high to low temperature following an {\em annealing schedule}.  After each temperature step the system is out of equilibrium and the equilibrating subroutine is used to move it back toward equilibrium. However, at low temperatures, the equilibrating subroutine is unable to facilitate transitions between different minima of the free energy landscape, and simulated annealing falls out of equilibrium if the weights associated with the free energy minima vary with temperature, as is typically the case. Thus simulated annealing cannot be used to sample equilibrium ensembles, and its primary use is to find ground states.  Population annealing solves this problem by simultaneously cooling a population of replicas of the system through a sequence of temperatures.  Each temperature step is associated with a resampling of the population so that some replicas are copied and other replicas are destroyed in such a way that the replicas are correctly weighted in the colder ensemble. In this way, at least for large populations, the population remains close to equilibrium as the system is cooled.  The resampling step in population annealing is similar to methods used in diffusion Monte Carlo~\cite{Anderson75} and the ``go with the winner" strategy~\cite{Grass2002}.

Sequential Monte Carlo methods~\cite{DoFrGo01}, of which population annealing is an example, are not well known in statistical physics but have been widely applied in statistics and finance.  One purpose of this paper is to bring this general method to the attention of computational statistical physicists.  We argue that PA may have an important role to play in simulations of systems with rough free energy landscapes and has some advantages over parallel tempering,  especially in situations where a moderately accurate result is required quickly and parallel computing resources are available.
 
 The outline of the paper is as follows.  We describe population annealing in Sec.\ \ref{sec:pa} and parallel tempering in Sec.\ \ref{sec:pt}.  Section \ref{sec:dw} introduces the two-well free energy landscape, and Sec.\ \ref{sec:padw} analyzes the performance of population annealing in this landscape.  Section \ref{sec:pavpt} compares the performance of population annealing and parallel tempering, and the conclusions of this section are supported by numerical results presented in Sec.\ \ref{sec:nr}.  Section \ref{sec:disc} concludes the paper with a discussion.

\section{Population Annealing}
\label{sec:pa}
The population annealing algorithm operates on a population of $R$ replicas of the system.  For disordered spin systems, each replica has the same set of couplings.  The algorithm consists of cooling the population of replicas through an annealing schedule from a high temperature, where equilibrium states are easily sampled, to a low target temperature, where the equilibrating subroutine cannot by itself feasibly equilibrate the system.   The annealing schedule is defined by a set of $\T+1$ inverse temperatures, 
\begin{equation}
\label{eq:repbb}
\bb_0 > \bb_1> \ldots, > \bb_{\T}.
\end{equation}
 The highest temperature, $1/\bb_{\T}$, is chosen to be a temperature for which it is easy to equilibrate the system.  It is often convenient to choose $\bb_{\T}=0$ as this facilitates the calculation of the absolute free energy of the system at each temperature in the annealing schedule.
  
In each temperature step the population is resampled and the equilibrating subroutine is applied to every replica at the new temperature.  The first part of a temperature step from $\bb$ to $\bp$ is resampling the population so that lower energy replicas are multiplied while higher energy replicas are eliminated from the population.   Suppose that the population is in equilibrium at $\bb$;  the relative weight of a replica $\g$ with energy $E_\g$ at inverse temperature $\bp$ is given by $\exp\left[ -(\bp-\bb)E_\g\right]$.
Thus, the expected number of copies of replica $\g$ that appear in the resampled population at $\bp$ is  
\begin{equation}
\label{eq:wn}
\wn_\g(\bb,\bp)=\frac{\exp\left[-(\bp-\bb)E_{\g}\right]}{\wt(\bb,\bp)},
\end{equation}
where $\wt$ is the normalization given by
\begin{equation}
\label{eq:Q}
\wt(\bb,\bp)=\frac{\sum_{j=1}^{R} \exp\left[-(\bp-\bb)E_{\g}\right]}{R} .
\end{equation}

The new population of replicas is generated by resampling the original population such that the expected number of copies of replica $\g$ is $\wn_\g$.  The actual number of copies $n_1,n_2,\ldots,n_R$ of each replica in the new population is given by the multinomial distribution $p\left[R;  n_1,\dots,n_R;  \wn_1/R, \ldots,  \wn_R/R \right]$ for $R$ trials.    In this implementation the population size is fixed. Other valid resampling methods are available.  For example, the number of copies of replica $\g$ can be chosen as a Poisson random variable with mean proportional to $\wn_\g(\beta,\beta^\prime)$, in which case the population size fluctuates~\cite{Mac10a}.  

For large $R$ and small $(\bp-\bb)$, the resampled distribution is close to an equilibrium ensemble at the new temperature, $\bp$.  However,  the regions of the equilibrium distribution for $\bp$ that differ significantly from the equilibrium distribution for $\bb$ are not well sampled, leading to biases in the population at $\bp$.  In addition, due to resampling, the replicas are no longer independent.  To mitigate both of these problem, the equilibrating subroutine is now applied.  Finally, observables are measured by averaging over the population.  

The entire algorithm consists of $\T$ steps: in step $k$ the temperature is lowered from $\bb_{\T-k+1}$ to $\bb_{\T-k}$ via resampling followed by the application of the equilibrating subroutine and data collection at temperature $\bb_{\T-k}$.

Population annealing permits one to estimate free energy differences. If the annealing schedule begins at infinite temperature corresponding to $\beta_{\T}=0$, then it yields an estimate of the absolute free energy $\F(\bb_k)$ at every temperature in the annealing schedule.  The following calculation shows that the normalization factor $\wt(\beta,\beta^\prime)$ is an estimator of the ratio of the partition functions at the two temperatures:
\begin{eqnarray}
\label{eq:ratioz}
\frac{Z(\beta^\prime)}{Z(\beta)} &=& \frac{\sum_\gamma e^{-\beta^\prime E_\gamma}}{Z(\beta)}\nonumber\\
&=& \sum_\gamma e^{-(\beta^\prime-\beta) E_\gamma} ( \frac{e^{-\beta E_\gamma}}{Z(\beta)} )
\nonumber\\
&=& \langle e^{-(\beta^\prime-\beta) E_\gamma} \rangle_\beta
\nonumber\\
&\approx&  \frac{1}{R} \sum_{j=1}^{R} e^{-(\beta^\prime-\beta) E_{\g}}=\wt(\beta,\beta^\prime) .
\end{eqnarray}
The summation over $\gamma$ is a sum over the microstates of the system while the sum over $j$ is a sum over the population of replicas in PA.  The last approximate equality becomes exact in the limit $R \rightarrow \infty$.  From Eq.\ \ref{eq:ratioz}  the estimated free energy difference from $\bb$ to $\bp$ is found to be
\begin{equation}
\label{eq:freediff}
-\bp \F(\bp) = -\bb F(\bb) +  \log \wt(\bb,\bp) ,
\end{equation}
where $F(\bb)$ is the free energy at $\bb$ and $\F$ is the estimated free energy at $\bp$. Given these free energy differences, if $\bb_{\T}=0$, then the PA estimator of the absolute free energy at each simulated temperature  is
\begin{equation}
\label{eq:sumQ}
-\beta_k \F(\beta_k) = \sum_{\ell=k+1}^{\T} \log \wt(\beta_{\ell},\beta_{\ell-1}) + \log \Omega ,
\end{equation}
where $\Omega=\sum_\gamma 1$ is the total number of microstates of the system; i.e.\ , $k_B \log \Omega$ is the infinite temperature entropy.

\section{Parallel Tempering}
\label{sec:pt}

Parallel tempering, also known as replica exchange Monte Carlo, simultaneously equilibrates a set of $R$ replicas of a system at $\T$ inverse temperatures 
\begin{equation}
\label{eq:repbb}
\bb_0 > \bb_1> \ldots, \bb_{\T-1} .
\end{equation}
There is one replica at each temperature so that $R=\T$ in contrast to population annealing, where typically the number $R$ of replicas greatly exceeds the number $\T$ of temperatures; i.e.\ , $R\gg \T$.
The equilibrating subroutine operates on each replica at its respective temperature. Replica exchange moves are implemented that allow replicas to diffuse in temperature space.  The first step in a replica exchange move is to propose a pair of replicas $(k,k-1)$ at neighboring temperatures $\bb=\bb_k$ and $\bp=\bb_{k-1}$.  The probability for accepting the replica exchange move is
\begin{equation}
\label{eq:reprob}
\ps=\min\left[1, e^{(\bb-\bp)(E-E^\prime)}\right].
\end{equation}
Here $E$ and $E^\prime$ are the respective energies of the replicas that were originally at $\bb$ and $\bb^\prime$.  If the move is accepted, the replica equilibrating at $\bb$ is now set to equilibrate at $\bp$ and vice versa.   Equation \ref{eq:reprob} insures detailed balance so that the Markov chain defined by parallel tempering converges to a joint distribution whose marginals are equilibrium distributions at the $\T$ temperatures of Eq.\ \ref{eq:repbb}.  Diffusion of replicas in temperature space allows round trips from low to high temperature and back. The benefit of these roundtrips is that free energy barriers are crossed in a time that grows as a power of the barrier height~\cite{Mac09a} rather than exponentially with respect to the barrier height as is the case for most single temperature dynamics.  Optimization schemes for PT depend in part on adjusting parameters to maximize the rate of making roundtrips~\cite{KaTrHuTr06,TrTrHa06, BiNuJa08}.

\section{Two-Well Model Free Energy Landscape}
\label{sec:dw}
In this section we describe a simple free energy landscape with two minima such as occurs, for example, in the low temperature phase of the Ising model or $\phi^4$ field theories.  This free energy was introduced in Ref.\ \cite{Mac09a} in the context of analyzing the efficiency of parallel tempering.   For $\bb \geq \bc$ and $\bc$ a critical temperature, the free energy $F_\s(\bb)$ associated with each well is defined by
\begin{equation}
\label{eq:F}
\bb F_\s(\bb) = -\frac{1}{2}(\bb-\bc)^2 \sig_\s  ,
\end{equation}
where
\begin{equation}
\label{eq:diste}
\sig^2_\s = \left\{ 
\begin{array}{l l}
  \N + H/2 & \quad \mbox{if $\s=1$}\\
  \N - H/2 & \quad \mbox{if $\s=0$}\\ \end{array} \right.
  \end{equation}
and $\s$ labels the well. The deep well corresponds to $\s=1$ and the shallow well to $\s=0$.  The well index $\s$ is the only macroscopic parameter in the model and the ``landscape'' is zero dimensional.  However, we also assume that the free energy at the saddle point between the wells is zero so that  $F$ is the free energy barrier between the wells.  The landscape is flat at $\bb=\bc$.  The parameter $\N$ is a proxy for system size.  In more realistic systems, barrier heights typically grow as a power of the number of degrees of freedom $N$ of the system.

The statistics of the energy of microstates in each well follows from this free energy using thermodynamics. The internal energy $U_\s(\bb)$ is the average of the energy distribution in well $\s$ and is obtained from
\begin{equation}
\label{eq:U}
U_\s(\bb)= \frac{\partial \bb F_\s}{\partial \bb} = -(\bb-\bc) \sig^2_\s.
\end{equation}
Using the relationship between specific heat and energy fluctuations, we find that  the variance  of the energy in well $\s$ is simply $\sig^2_\s$.

The free energy also determines the probability $\pp(\bb)$ of being in the deep well according to
 \begin{equation}
\label{eq:pp}
\pp(\bb)={\bf E}(\s)=\frac{1}{1+e^{- (\bb-\bc)^2 H}} .
\end{equation}
Microstates of the model are specified by an energy and a well index. We assume that in equilibrium, the distribution of energies, conditioned on the well index $\s$, is a normal distribution with mean $U_\s(\bb)$ and variance $\sig^2_\s$.  Thus, the well index $\s$ is a Bernoulli random variable such that $\s=1$ with probability $\pp(\bb)$ and  $\s=0$ with probability $1-\pp(\bb)$. The energy $E$ is given by
\begin{equation}
\label{eq:diste}
E = \left\{ 
\begin{array}{l l}
  N(U_1(\bb),\sig_1^2) & \quad \mbox{if $\s=1$},\\ 
 N(U_0(\bb),\sig_0^2) & \quad \mbox{if $\s=0$},\\  \end{array} \right.
  \end{equation}
 where $N(\mu,\Delta^2)$ is a normal random variable with mean $\mu$ and variance $\Delta^2$.
 
The dynamics of the model under the equilibrating subroutine for $\bb \leq \bc$ is assumed to have the following properties.  The well index is conserved except at the critical temperature,  $\bc$.  That is, there are no transitions between the wells except for $\bb=\bc$.  On the other hand, for $\beta>\bc$ the equilibrating subroutine is assumed to equilibrate the system within each well in a single time unit.  Thus, the sequence of energies produced by successive steps of the equilibrating subroutine will be i.i.d.\ normal random variables  $N(U_\s(\bb),\sig_\s^2)$, where $\s$ is the well index.  For $\bb=\bc$ the equilibrating subroutine first chooses one of the wells with equal probability and then chooses the energy from the associated normal distribution.

\section{Convergence of Population Annealing}
\label{sec:padw}

We  first consider a single step of population annealing from inverse temperature $\bb$ to inverse temperature $\bb^\prime > \bb$.  We will compute the error made by population annealing in the free energy and the fraction of the population in the deep well as a function of the number of replicas, the size of the temperature step $\bb^\prime- \bb$, and the parameters $\bc$, $K$ and $H$ of the two-well model.  Let 
\begin{equation}
Y_\g=\exp\left[ -(\beta^\prime-\beta)E_\g  + \lambda \sr_\g\right] ,
\end{equation}
where $E_\g$ is the energy and $\sr_j$ is the well index of replica $\g$.  Setting $\lambda=0$ yields the un-normalized re-weighting factor (see Eq.\ \ref{eq:wn}) of replica $\g$ from inverse temperature $\bb$ to $\bb^\prime$.  The extra term $ \lambda \sr_\g$ in the exponent will be used to calculate the probability of being in the deep well.  

To obtain the error in the free energy and the fraction of replicas in the deep well at temperature $\bb^\prime$ assuming the correct equilibrium distribution at $\bb$, we compute $I(\bb,\bb^\prime,\lambda)$, the expectation of the logarithm of $Y$:
\begin{equation}
I(\bb,\bb^\prime,\lambda)={\bf E}\log (\frac{1}{R}\sum_{j=1}^R Y_j) .
\end{equation}
Using Eq.\ \ref{eq:ratioz}, we obtain the PA estimate of the free energy difference by setting $\lambda=0$ in $I(\bb,\bb^\prime,\lambda)$:
\begin{equation}
\label{ eq:bf}
\bp \F(\bp)-\bb F(\bb) = -I(\bb,\bb^\prime,0) \equiv {\bf E}( \log \wt(\bp,\bb)) .
\end{equation}
In this equation, $\bb^\prime \F(\bb^\prime)$ is the estimate of the free energy at $\bp$ given the exact value at $\bb$.  The fraction in the deep well at the lower temperature is obtained by differentiation with respect to $\lambda$:
\begin{equation}
\label{eq:cprime}
\ppt(\bb^\prime)= \frac{dI(\bb,\bb^\prime,\lambda)}{d\lambda}\bigg|_{\lambda=0}.
\end{equation}
Our goal is to determine how much these estimates deviate from the corresponding exact values.

Let $S_n$ be a sum of $n$ independent, identically distributed random variables $X_j$.
Generically, one can use Taylor's Theorem to prove that the leading terms in an asymptotic expansion of the expectation of a function $f(S_n/n)$ have the form
\begin{equation}
\label{eq:efy}
{\bf E}f(S_n/n) ={\bf E}f(\sum_{j=1}^n X_j/n)= f({\bf E} X) + \frac{1}{2 n} f^{\prime\prime}({\bf E} X)  {\bf Var } X +\mbox{O}\!\left(\frac{1}{n^{3/2}}\right) \!,
\end{equation}
where ${\bf E}(X)$ and ${\bf Var }(X)$ denote the expectation and variance of $X_j$, respectively.  
Thus, for our case,
\begin{equation}
\label{eq:i}
I(\bb,\bb^\prime,\lambda) = \log({\bf E} Y) - \frac{1}{2 R}   \frac{{\bf Var } Y}{({\bf E} Y)^2} +\mbox{O}\!\left(\frac{1}{R^{3/2}}\right) \! .
\end{equation}
The first term is the exact result and the second term is the leading order systematic error in the population annealing estimate due to a finite population size.  Setting $\lambda=0$, we have
\begin{equation}
\label{eq:ef}
\bp \F(\bp)-\bp F(\bp)= \frac{1}{2 R}   \frac{{\bf Var } Y}{({\bf E} Y)^2} +\mbox{O}\!\left(\frac{1}{R^{3/2}}\right) \!
\end{equation}
This result shows that the systematic error decreases as the inverse of the population size and that the free energy approaches the exact value from above as the number of replicas increases. 

The variance of the free energy estimator was observed to be a useful measure of the convergence of the algorithm~\cite{Mac10a}.  Here we formalize that observation by computing the variance of $\bp \F(\bp)$ presuming that  $\bb F(\bb)$ is exactly known and thus has no variance.  The variance of the free energy estimator is given by, 
\begin{equation}
\label{eq:varff}
{\bf Var } (\bp \F(\bp)) = {\bf E} \left( \log^2 \frac{1}{R} \sum_{j=1}^{R} e^{-(\beta^\prime-\beta) E_{\g}} \right) - \left( {\bf E} \log \frac{1}{R} \sum_{j=1}^{R} e^{-(\beta^\prime-\beta) E_{\g}}\right)^2 .
\end{equation}
Applying Eq.\ \ref{eq:efy} to both $\log$ and $\log^2$ and substituting the results into Eq.\ \ref{eq:varff} yields
\begin{equation}
\label{eq:varf}
{\bf Var } (\bp \F(\bp) )=  \frac{1}{ R}   \frac{{\bf Var } Y}{({\bf E} Y)^2} +\mbox{O}\!\left(\frac{1}{R^{3/2}}\right) \! ,
\end{equation}
where $\lambda$ is set to zero on the RHS of this equation.
Comparing Eqs.\ \ref{eq:varf} and \ref{eq:ef}, we find that
\begin{equation}
\label{eq:varef}
\bp \F(\bp)-\bp F(\bp) = \frac{1}{2}{\bf Var } (\bp \F(\bp) )+\mbox{O}\!\left(\frac{1}{R^{3/2}}\right) \!  .
\end{equation}
This equation is useful because ${\bf Var } (\bp \F(\bp) )$ can be directly estimated from multiple runs of PA, and thus the accuracy of the algorithm as applied to a specific system can be estimated from the algorithm itself.  Although Eq.\ \ref{eq:varef} was derived for a single step of PA for  the two-well model, the calculation does not rely on specific features of the model.  Furthermore, since the variance is additive  we conjecture that Eq.\ \ref{eq:varef} is a good approximation for the full PA algorithm applied to any statistical mechanical system.  In support of this conjecture, we note that Eq.\ \ref{eq:varef} is a good approximation for the PA estimate of the low temperature free energy of the one-dimensional Ising spin glass studied in \cite{Mac09a} for which the exact free energy can be calculated using transfer matrix methods.

For the case of the two-well model, we can evaluate the Gaussian integrals exactly for both the mean and variance of the weight factor $Y$.  The general result is 
\begin{eqnarray}
\label{eq:fulli}
I(\bb,\bb^\prime,\lambda) & = & \frac{\N}{2} \left[(\bb^\prime-\bc)^2-(\bb-\bc)^2\right]  \\
&& + \log \cosh\left[(\bb^\prime-\bc)^2 \frac{H}{4}+\lambda/2\right] 
- \log \cosh\left[(\bb-\bc)^2 \frac{H}{4}\right] +\lambda/2 \nonumber   \\
&&+ \frac{1}{2R}-\frac{1}{2R}\exp\left[ (\beta^\prime-\beta)^2\N\right]\Big( \frac{\cosh\left[(\bb-\bc)^2 H/4\right]\cosh\left[(2\bb^\prime-\bb-\bc)^2 H/4 +\lambda\right] }{\cosh\left[(\bb^\prime-\bc)^2 H/4 +\lambda/2\right]^2 }\Big) \nonumber \\ 
&&+\mbox{O}\!\left(\frac{1}{R^{3/2}}\right) \! \nonumber  .
\end{eqnarray}
From this general result it is instructive to consider expansions to first order in $H$.  The estimate of the free energy difference is
 \begin{eqnarray}
\label{eq:freestep}
\bp \F(\bp)-\bb F(\bb) &=& -\frac{\N}{2} \left[(\bp-\bc)^2-(\bb-\bc)^2\right] \\ \nonumber
 &&+  \frac{1}{2R}\left[\exp((\bp-\bb)^2 \N)-1\right] +\mbox{O}\!\left(\frac{1}{R^{3/2}}\right) \! +\mbox{O}\!\left(H^2\right).
\end{eqnarray}
The first term on the RHS of this expression is the exact free energy difference in the two-well model for symmetric wells.  The second term is the error made by population annealing, which decreases inversely in $R$.  The form of the error term also reveals that the size of the temperature steps should be  $(\bp-\bb) \lesssim 1/\sqrt{\N}$ to keep the error under control as the barrier height increases.

The error in the free energy estimate at $\bb_0$ for small $H$ can be obtained by summing the errors made in each temperature step of the algorithm.  Since the errors depend only on the size of the temperature step, the algorithm is optimized with constant size steps in $\bb$.  Thus the error estimate at $\bb_0$ is
 \begin{equation}
\label{eq:freetotal}
\bb_0 \F(\bb_0)-\bb_0 F(\bb_0) = \frac{\T}{2R}\left(\exp\left[(\bb_0-\bc)^2 \N/\T^2\right]-1\right)+\mbox{O}\!\left(\frac{1}{R^{3/2}}\right) \!  +\mbox{O}\!\left(H^2\right).
\end{equation}

Next we consider the fraction of the population in the deep well at temperature $\bb^\prime$ given the equilibrium value at $\bb$. Combining Eq.\ \ref{eq:cprime} and \ref{eq:fulli} and expanding to leading order in $H$, we obtain
\begin{eqnarray}
\label{eq:cr}
\ppt(\bb^\prime) &= &\frac{1}{2} +\frac{H (\bb^\prime-\bc)^2}{8} \\  \nonumber
&&-\frac{H}{8R} (\beta^\prime-\beta)(3\beta^\prime-\beta-2\beta_c) \exp\left[(\beta^\prime-\beta)^2 \N \right] +\mbox{O}\!\left(\frac{1}{R^{3/2}}\right)+\mbox{O}\!\left(H^2\right).
\end{eqnarray}
The first two terms on the RHS of this expression are the leading order in $H$ expansion of the exact value, Eq.\ \ref{eq:pp}.  The correction term shows that again the temperature steps should satisfy  $(\bp-\bb) \lesssim 1/\sqrt{\N}$.   

From Eq.\ \ref{eq:cr} we could obtain an estimate of the overall error in the fraction in the deep well by summing over the $\T$ temperature steps.  Unfortunately the result significantly underestimates the true error.  The reason is that the resampling step introduces correlations between replicas so that the probability distribution for the number of replicas in the deep well has a variance that is broader than that of the binomial distribution assumed in the above analysis.  Nonetheless, we conjecture that the leading term in the error made by the full algorithm in the fraction in the deep well, $(\ppt(\bb_0) -\pp(\bb_0))$, behaves as $1/R$ and can be minimized when  $\T \sim \sqrt{\N}$.  The effect of correlations is much less important for the free energy, as evidenced by the absence of a term that is order $H$ in Eq.\ \ref{eq:freestep}, and we  conjecture that Eq.\ \ref{eq:freetotal} is exact to leading order.  We are currently studying these questions. 

The two main conclusion from this analysis are that (1) the error decreases inversely with the number of replicas and (2) the error can be made small only if the temperature step size satisfies $(\bp-\bb) \lesssim 1/\sqrt{\N}$.

\section{Parallel Tempering vs Population Annealing} 
\label{sec:pavpt}
How do PT and PA compare in the efficiency with which they converge to equilibrium?   Here we estimate the amount of computational work needed to make the deviation from equilibrium small.  The quantities that we increase, holding other parameters fixed, are the number of replicas $R$ for PA and the number of sweeps $t$ for PT.   Within the stylized two-well model, we define computational work $\W$ as the total number of times replicas are acted on by the equilibrating subroutine.  For PA, with $R$ replicas and $\T$ temperature steps the work is $\W=R\T$.   For  PT with $R$ replicas the computational work  is given by $\W=Rt$ where $t$ is the number of PT sweeps.  

Our measure of computational work ignores the time required to resample the population in PA or implement replica exchange in PT.  For large systems, this time is negligible compared to the time spent in the equilibrating subroutine.  The quantity $\W$ assigns one unit of time for one sweep of the equilibrating subroutine for a single replica; thus the computational work measured in elementary operations rather than sweeps of the equilibrating subroutine is $N\W$, where $N$ is the number of degrees of freedom of the system.  Since $N$ is the same for PA and PT and not explicitly defined in the two-well model, we don't consider this factor explicitly. 

Suppose we carry out the simulations on a massively parallel computer and consider parallel time instead of sequential time (work).  Since each replica can be independently acted on by the equilibrating subroutine, one parallel time unit is required for one sweep of all the replicas.  Thus, for the highly parallel PA, the parallel time is the number of temperature steps $\T$ whereas for the less parallel PT, the parallel time is the number of PT sweeps $t$.  

In \cite{Mac09a} we analyzed the efficiency of PT for the two-well model.  As is generally the case for Markov chain Monte Carlo methods, convergence to equilibrium is asymptotically exponential.  The deviation from equilibrium is controlled by an exponential autocorrelation time $\te$, and its leading behavior is proportional to $\exp(-t/\te)$, where $t$ is the number of Monte Carlo sweeps.  In the two-well model, for small asymmetries between the wells (i.e., $(\bb_0-\bc)^2 H \leq 1$), $\te$ is controlled by the diffusive time scale for a replica to diffuse between the lowest and highest temperature.  If the number of temperatures is sufficiently large, then the acceptance fraction for replica exchange is not small and the elementary time step in this diffusive process is order unity so that $\te \sim R^2$.  The optimum number of replicas was shown to scale as $R=\T+1 \sim \sqrt{\N}$ and given this choice, $\te \sim R^2 \sim \N$.   When the asymmetry becomes large, the optimum number of temperatures remains approximately the same but there is a crossover to a ballistic regime and $\te \sim R \sim  \sqrt{\N}$.   Convergence to equilibrium occurs on a time scale $\te$ so that, in the small-asymmetry diffusive regime, the work $\W_0$ required to begin to achieve moderately accurate results is given by $\W_0 \sim R\te \sim \N^{3/2}$. 

For PA we found in Sec.\ \ref{sec:padw}, Eq.\ \ref{eq:fulli} that the error term depends on $\N$ as  $\exp\left[(\beta^\prime-\beta)^2 \N \right]$ so that the optimum number of temperature steps scales as $\T \sim  \sqrt{\N}$, just as is the case for PT.  Based on the form of Eq.\ \ref{eq:cr}, we conjecture that the overall error behaves as $S^a/R$ with $a$ an exponent less than or equal to unity. Thus, the error decreases as $\T^{1+a}/\W \sim \N^{(1+a)/2}/\W$, and the computational work $\W_0$ required to begin to achieve moderately accurate results behaves as  $\W_0 \sim  \N^{(1+a)/2}$.  For large systems (large $\N$) and nearly degenerate free energy minima, population annealing is expected to be more efficient initially than parallel tempering by a factor of a power of the barrier height, $\N^{1-a/2}$.  However, for large amounts of computational work (i.e.,  $\W \gg R\te \sim \N^{3/2}$)  PT is much closer to equilibrium than PA because PT converges exponentially in $t$ while PA converges inversely in the comparable variable $R$.

\section{Numerical Results}
\label{sec:nr}
We have carried out simulations of the two-well model using both PT and PA to compare the efficiency of the algorithms and test the conjectures of the previous section. In these simulations, the equilibrating subroutine samples a Gaussian random number with mean and variance appropriate to the temperature and well-index of the replica, Eq.\ \ref{eq:diste}.  The well index is a conserved quantity except at $\bc$. At $\bc$ the equilibrating subroutine first chooses the well index with equal probability and then chooses the energy according to Eq.\ \ref{eq:diste} with $\bb=\bc$ and the given value of $\s$. 
 
\begin{figure}
\centering
\subfigure[$\N=16$, $S=11$]{\label{fig:16}\includegraphics[width=3.5in]{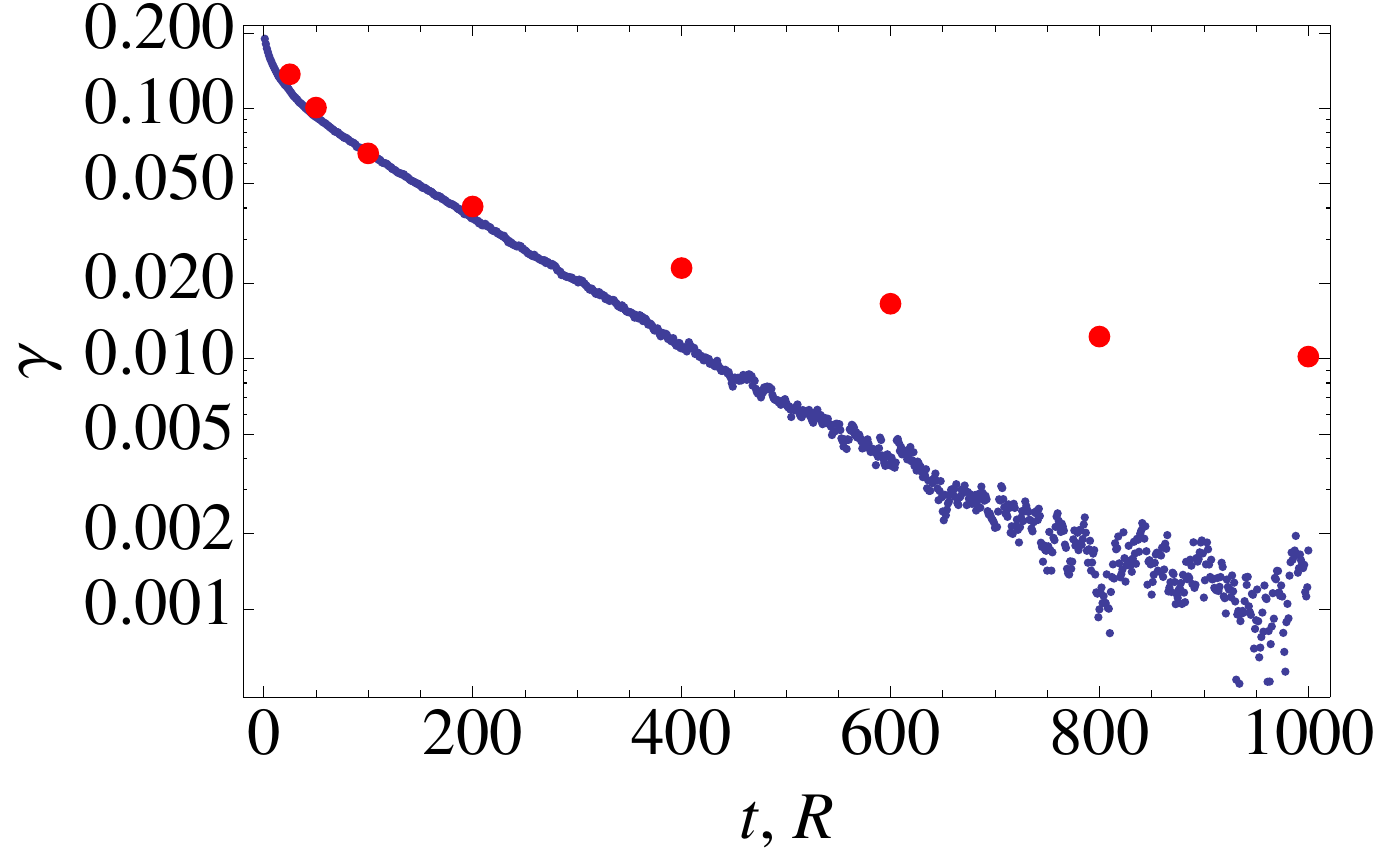}}
\subfigure[$\N=64$, $S=23$]{\label{fig:64}\includegraphics[width=3.5in]{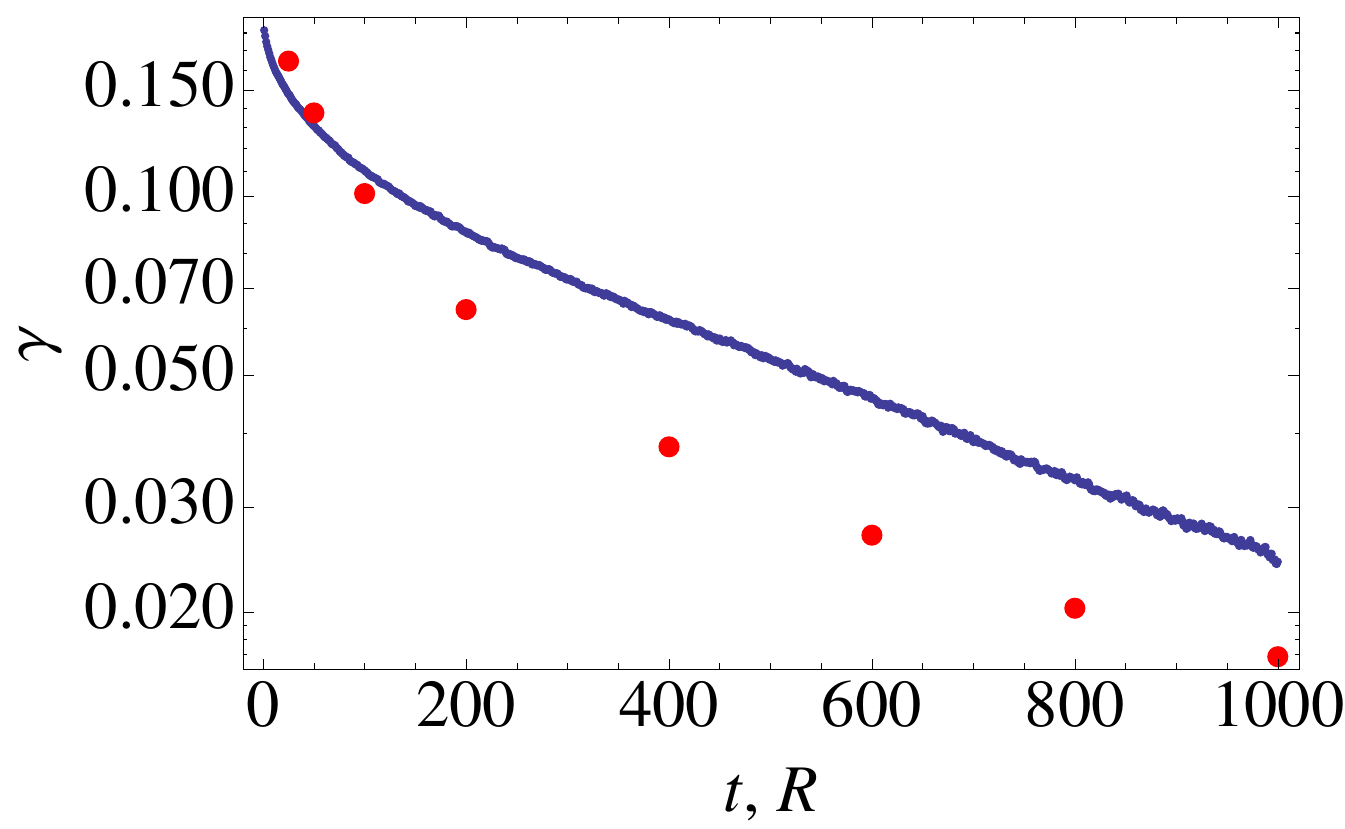}}
\subfigure[$\N=256$, $S=47$]{\label{fig:256}\includegraphics[width=3.5in]{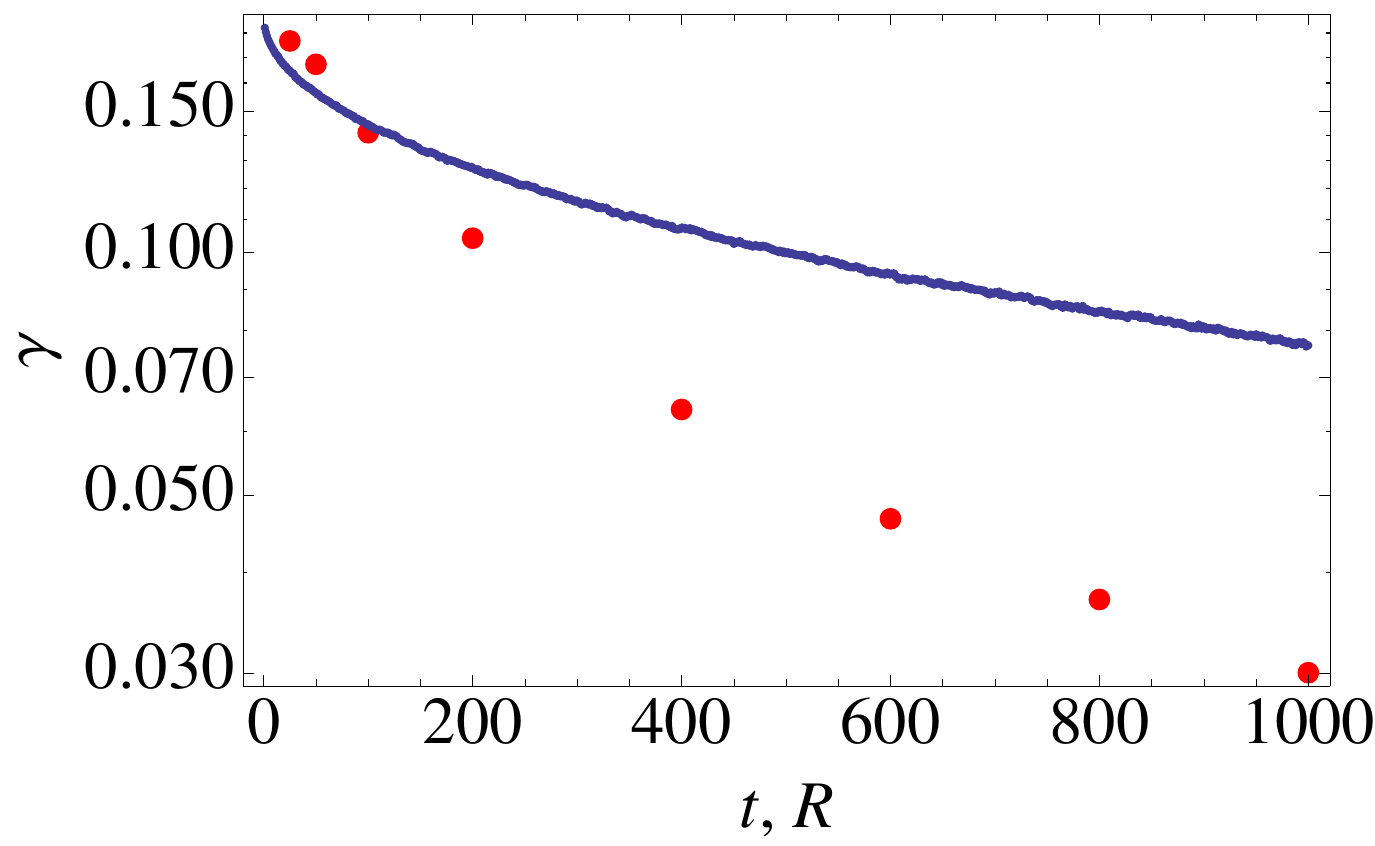}} 
\caption{The deviation of the fraction in the deep well from the equilibrium value $\gamma$ vs.\ Monte Carlo sweeps $t$ for PT (small blue points) or population size $R$ for PA (large red points) for the values of $\N$ and $\T$ shown in the subcaptions.  For each case $H=0.1$. }     
\end{figure}

Figures \ref{fig:16}, \ref{fig:64} and  \ref{fig:256} show the convergence to equilibrium of $\gamma$, the deviation from equilibrium of the probability of being in the deep well at the lowest temperature given by
\begin{equation}
\label{ }
\gamma=\ppt(\bb_0)-\pp(\bb_0),
\end{equation}
as a function the number of sweeps $t$ for  PT (small blue points) or populations size $R$ for PA (large red points).  The horizontal axis therefore measures computational work for both algorithms in the same units.  The figures differ according to the value of the well depth $\N$ and number of temperatures $\T$ with $\N=16$, $\T=11$ for Fig.\ \ref{fig:16}, $\N=64$, $\T=23$ for Fig.\ \ref{fig:64} and $\N=256$, $\T=47$ for Fig.\ \ref{fig:256}.   In each case, there is a small asymmetry, $H=0.1$ and at the lowest temperature $\pp(\bb_0)=0.68997$.  The highest and lowest temperatures are $\bc=1$ and $\bb_0=5$, respectively, and the number of temperatures  $\T+1$ is chosen to be close to the optimum value for PT and to scale as $\sqrt{K}$.  
The simulations confirm the conclusions of Sec.\ \ref{sec:pavpt} and show that for larger systems (larger $\N$) PA is initially closer to equilibrium than PT for the same amount of computational work.
%figures produced by the mathematica file popan.nb

\begin{figure}
 \includegraphics[width=5in]{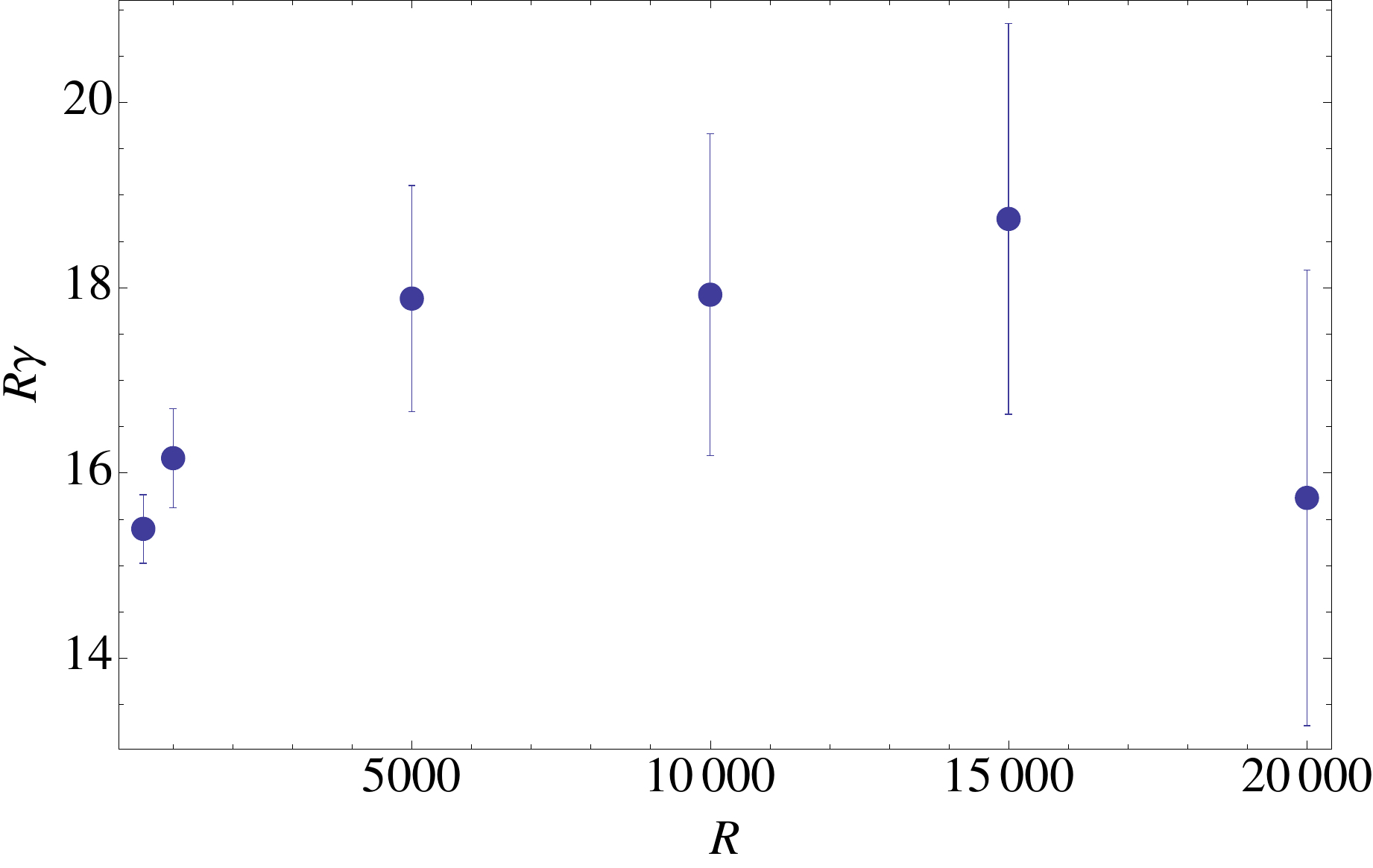}
\caption{ $R\gamma$  vs.\ $R$ for $\N=64$, $\T=23$ and $H=0.1$.}
\label{fig:rgvr}       
\end{figure}

We have also tested two conjectures concerning the convergence of PA for the two-well model.  The first conjecture is that the error in the fraction in the deep well decreases inversely in the population size.  Figure \ref{fig:rgvr} shows $R\gamma$ as a function of $R$ for the case $H=0.1$, $\N=64$, and $\T=23$. 
 It is clear that within the error bars $\gamma$ is behaving as $1/R$ over the range of $R$ studied.  The averages and error bars are obtained from 10$^5$ independent runs of PA for each population size.  
 
 The second conjecture is that for large $R$ and $\T \sim \sqrt{\N}$
\begin{equation}
\label{eq:gvk}
R\gamma \sim K^{a/2}.
\end{equation}
\begin{figure}
 \includegraphics[width=5in]{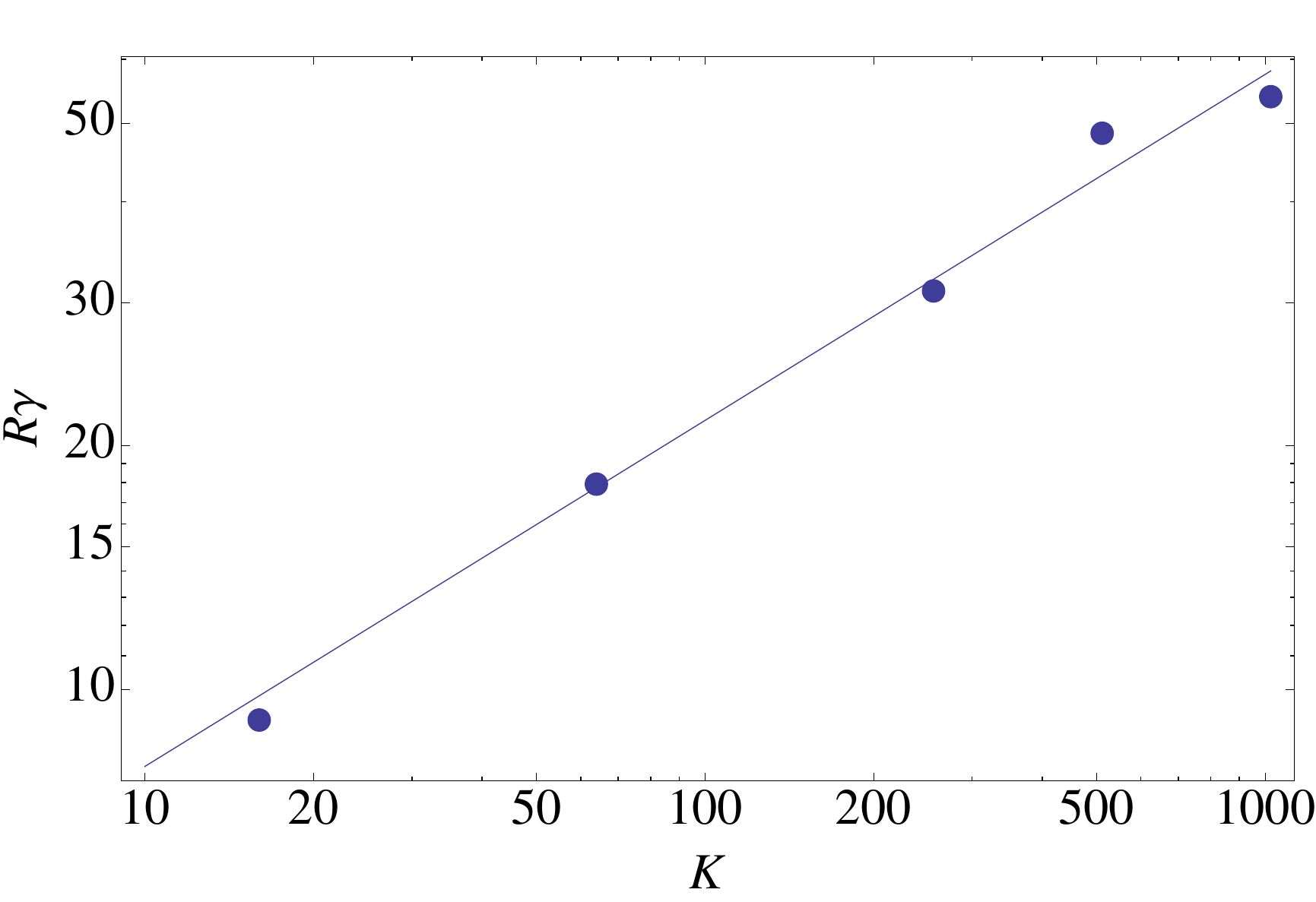}
\caption{ $R\gamma$ vs.\ $\N$ for $R=10000$ and $H=0.1$.  The solid line is the best power-law fit, $R\gamma = K^{0.43}$.}
\label{fig:rgvk}       
\end{figure}
Figure \ref{fig:rgvk} is a log-log plot of $R\gamma$  vs.\ $\N$.  For each value of $K$, PA is run with a large population size $R=10000$, $H=0.1$, and the values of $\T$ given above satisfying $\T \sim \sqrt{\N}$.  Averages and errors are obtained from 10$^5$ independent runs.  The best fit to Eq.\  \ref{eq:gvk} is shown as the solid line and yields $a=0.85 \pm 0.08$.  Note that $a\leq1$ as conjectured, supporting the hypothesis that for modest amounts of computational work, PA yields a smaller error in $\gamma$ than PT and that this advantage increases with well-depth parameter $\N$.

\section{Discussion}
\label{sec:disc}
We have seen that both parallel tempering and population annealing are able to solve  the problem of sampling equilibrium states of systems having several minima in the free energy landscape separated by high barriers.  Parallel tempering is a Markov chain Monte Carlo method and thus converges to equilibrium exponentially in the number of sweeps, whereas population annealing converges inversely in the population size.  Their relative efficiencies are described qualitatively by the parable of the tortoise and the hare.    For a given amount of computational work, the hare (population annealing) is initially closer to equilibrium  but ultimately the tortoise (parallel tempering) catches up and gets ahead.  

The problem of high barriers between different free energy minima is only one of the difficulties encountered in simulating systems with rough free energy landscapes.  A second, generic problem is that the relevant free energy minima may have small basins of attraction so that when the system is annealed starting from high temperature, it is very unlikely that the relevant low temperature states will be found.  Small basins of attraction occur for first order transitions and for NP-hard combinatorial optimization problems and are almost certainly a feature of the low temperature phase of spin glasses.  In this situation a very large number of sweeps of parallel tempering or a very large population size in population annealing are required simply to find the relevant states. 

Parallel tempering is widely used in several areas of computational physics while population annealing is not well known.  One of the conclusions of this paper is that population annealing is an attractive alternative to parallel tempering for studies where moderately accurate answers are required quickly.  This is especially the case if massively parallel computing resources are available since population annealing is well suited to parallelization.  

\begin{acknowledgments}
Jon Machta was supported in part from NSF grant DMR-0907235 and Richard Ellis by NSF grant DMS-0604071.

\end{acknowledgments}

\end{document}